\documentclass{article}
\usepackage{amsfonts}
\usepackage{amsmath}

\title{VRAC: Virtual Raw Anchor Coordinates Routing in Sensor Networks --
Concepts and Experimental Protocol}

\author{Florian Huc and Aubin Jarry}

\begin{document}
\maketitle

\subsection*{Abstract}

In order to make full use of geographic routing techniques developed for large scale networks, nodes must be localized. However, localization and virtual localization techniques in sensor networks are dependent either on expensive and sometimes unavailable hardware (e.g. GPS) or on sophisticated localization calculus (e.g. triangulation) which are both error-prone and with a costly overhead.

Instead of localizing nodes in a traditional 2-dimensional space, we intend to use directly the raw distance to a set of anchors to route messages in the multi-dimensional space. This should enable us to use any geographic routing protocol in a robust and efficient manner in a very large range of scenarios.

\section{Introduction, State of the Art}

The use of Sensor networks implies that they are made of cheap devices. Their objective is to make measurement on a wide area and to gather all this measures at one, or eventualy several sinks. One can also consider the scenari when any pair of sensor may want to communicate \cite{AI+02}.

If in wired networks where each node is equipped with huge computation and storage ressources, it is possible to maintain routing tables, these is not doable in sensor networks. Indead, the computation phase requires energy (which is a limited ressource) and the storage of the datas may be important in case of all to al communications. Instead of using routing tables, local routing techniques have been develloped. A compelling technique consists in using nodes's coordinates. One may suppose for example that each sensor is equipped with a GPS and so knows precisely it's position. This gives precious information to get closer to the destination by choosing greedily the closest neighbour. Howewer, one has to be carreful since obstacles may lead to local minimum before the destination is reached. So, routing algorithms have to detect obstacles. Many algorithms have been devised as OAFR \cite{KWZ08}, an extension of GPSR \cite{KK00} which consists in planarizing the connectivity graph and then doing face routing. One can also site GRIC \cite{PS07} a greedy routing following the sides of an obstacle when one is met, and which has some inertia in the direction followed by the message. If one authorized the use of a bit of memory at each node, then algorithm have been proposed to do early obstacle detection \cite{ML+08}.

However, one may argue that the hypothesis of having a GPS for each sensor leads to too expensive devices. Hence, we may want to weaken it, by, for exemple, equiping a subset of the sensors with GPS, these sensors are usualy called anchors. Then, these nodes which know their position will be exploited to compute approximate coordinates for all nodes of the network. These techniques needs flloding from anchors and many computation at each nods, hence they are energy conssuming. Furthermore, the computed coordinates are approximations which turns to be often insufficient. Some authors, improved these  results by using angles measurement \cite{BGJ09}. But here again, the coordinates may be insufficiently accurate, and such angles measurement need etra devices at each node whose cost needs to be compared to the one of a GPS.

It is also interesting to notice that technique to compute virtual coordonates without any anchor in the network exist  \cite{CC+05}. Such techniques are known under the name of global embedding. If they do not need any GPS, they suffer from inacurracy and energy consomption greediness.

In this paper, we consider that there are some special nodes in the network, to which any other node knows its distance. By similarity, we call them anchors. Our objective is to study routing techniques using directly the distance to the anchors as coordinates, without computing from them 2-dimensional coordinates. Our objective is to adapt this idea to any algorithm using greedy techniques, for a start, we do it with GRIC.

In the next section, we detail the idea and the forssen difficulties, then in the last section we detail the scenarios on which we will test our idea.

\section{Concept}

The multi-dimensional coordinates of a node at location $X$ is defined by its distances to some set of anchors at location $A_1, A_2, \dots A_n$:

$$f:X\rightarrow \begin{pmatrix} d(X,A_1) \\ d(X,A_2) \\ \dots \\ d(X,A_n) \end{pmatrix}.$$


In the plane with Euclidean distance, a node at location $X$ has absolute coordinates $(x,y)$, and anchors have absolute coordinates $(x_i, y_i)$ so $f$ is a function from $\mathbb{R}^2\rightarrow\mathbb{R}^n$ defined by
$$f:(x,y)\rightarrow \begin{pmatrix} \sqrt{(x-x_1)^2 + (y-y_1)^2} \\ \sqrt{(x-x_2)^2 + (y-y_2)^2} \\ \dots \\ \sqrt{(x-x_n)^2 + (y-y_n)^2} \end{pmatrix}.$$
Since the functions $f_i:(x,y)\rightarrow\sqrt{(x-x_i)^2 + (y-y_i)^2}$ are continuous and $C^\infty$ except in $(x_i,y_i)$, the image $f(\mathbb{R}^2)$ in $\mathbb{R}^n$ is a {\em continuous surface} with singularities at the image of anchors.
Observe that any continuous distance will produce a continuous surface in $\mathbb{R}^n$.


\subsection*{Foreseen difficulties}

\begin{itemize}
\item
The surface singularities near anchors is a consequence of the more general {\em anchor-in-the-middle} problem. Suppose that there are three locations $A, X, Y$ in the Euclidean plane such that $\vec{XY} = 3\times\vec{XA}$. While trying to route from point $X$ to point $Y$, a coordinate related to $A$ will tell $X$ to send the message away from $A$, since $X$ is closer to $A$ then $Y$, which is exactly the wrong thing to do in this context.
This problem is independent from the choice of the distance function, but is not without solutions. An escape solution would be to have only anchors on the boundary of the network. A second solution would be to have enough anchors so that the problem is mitigated. A third solution would be to wisely select anchors while routing, ignoring anchors whose distance to the sender is less than the distance to the destination and less than half the distance to the farthest anchor.

\item
The surface $f(\mathbb{R}^2)$ has a {\em curve}. If we use the Euclidean distance, the surface will be almost flat far away from the anchors, but the curve will be more pronounced near the anchors. The curve will probably lead to somewhat sub-optimal curved paths in the original $\mathbb{R}^2$. As an illustration, ships do not follow constant latitude paths, which may be up to $\frac{\pi}{2}$ times longer than wanted near the north or south pole.

\item
While saving on initialization overhead, multi-dimensional routing will cause some additional computation costs when sending messages. First, we point out that theses computation costs are not communication costs and should be lower in terms of energy consumption by some magnitude order. Then, we believe that the added costs will most likely be that of multi-dimensional scalar products, which are just some added additions and multiplications, while the expensive operations will stay the same as in traditional 2-dimensional routing.

\end{itemize}
\section{Scenarios}

We propose several scenari for which our technique should be used. There are two main types. In the first one, the anchors are special nodes of the network which can advertise their distance to the other nodes. We will first consider when such nodes are positionned on the boundary of the network, and then when they are chosen at random between all nodes of the network.

The second type of scenario is when the signal is emited by some external entities, as a plane or a robot which deploy the network. Just after it would deliver a powerfull signal at different positions which play the role of anchors. An other option is to have special devices, which can be seen as signal bombs, and whose only purpose is to emit a strong signal and to be an anchor.

\section{Experimental Protocol}

The algorithm on which we first test the technique is GRIC. The implementation is made using the simulator AlgoSensim. We compare the results with the traditionnal GRIC implemantation.

The first situation is when there is two anchors at infinite north/east (normal coordinates), it should give the same result.

Then we will test it with four to ten anchors, either on the boundaries or positionned at random. The distance to the anchors will be exact distance first and then the distance in terms of hop.

Finally, in the case of use of several anchors (10), we will consider the possibility of using the routing technique using only a subset of the coordinates, ie, considering only a reduced number of anchors from all available.

\label{sec:biblio}

\nocite{*}


\begin{thebibliography}{MLNR08}

\bibitem[ASSC02]{AI+02}
IF~Akyildiz, W.~Su, Y.~Sankarasubramaniam, and E.~Cayirci.
\newblock {Wireless sensor networks: a survey}.
\newblock {\em Computer networks}, 38(4):393--422, 2002.

\bibitem[BGJ09]{BGJ09}
J.~Bruck, J.~Gao, and A.A. Jiang.
\newblock {Localization and routing in sensor networks by local angle
  information}.
\newblock 2009.

\bibitem[CCDU05]{CC+05}
A.~Caruso, S.~Chessa, S.~De, and A.~Urpi.
\newblock {GPS free coordinate assignment and routing in wireless sensor
  networks}.
\newblock In {\em Proceedings IEEE INFOCOM 2005. 24th Annual Joint Conference
  of the IEEE Computer and Communications Societies}, volume~1, 2005.

\bibitem[KK00]{KK00}
B.~Karp and HT~Kung.
\newblock {GPSR: greedy perimeter stateless routing for wireless networks}.
\newblock In {\em Proceedings of the 6th annual international conference on
  Mobile computing and networking}, pages 243--254. ACM New York, NY, USA,
  2000.

\bibitem[KWZ08]{KWZ08}
F.~Kuhn, R.~Wattenhofer, and A.~Zollinger.
\newblock {An algorithmic approach to geographic routing in ad hoc and sensor
  networks}.
\newblock {\em IEEE/ACM Transactions on Networking}, 16(1):51--62, 2008.

\bibitem[MLNR08]{ML+08}
L.~Moraru, P.~Leone, S.~Nikoletseas, and J.~Rolim.
\newblock {Geographic Routing with Early Obstacles Detection and Avoidance in
  Dense Wireless Sensor Networks}.
\newblock {\em Lecture Notes in Computer Science}, 5198:148--161, 2008.

\bibitem[PN07]{PS07}
O.~Powell and S.~Nikoletseas.
\newblock {Simple and efficient geographic routing around obstacles for
  wireless sensor networks}.
\newblock {\em Lecture Notes in Computer Science}, 4525:161, 2007.

\end{thebibliography}
\end{document}